# Unmanned F/A-18 Aircraft Landing Control on Aircraft Carrier in Adverse Conditions

Mikhail Kistyarev [1], Xinhua Wang [1†]

1 Aerospace Engineering, University of Nottingham, UK

† Email: wangxinhua04@gmail.com

## Abstract

Carrier landing of aircrafts is a challenge for control due to the existence of nonlinear wind disturbances and the requirements of changing reference trajectories. In this paper, a robust landing control system is presented for carrier landing of unmanned F/A-18 aircraft. In the control system, an augmented observer is applied to estimate the combined disturbances in the pitch dynamics of F/A-18 aircraft during carrier landing. Therefore, the control performance is improved through the control compensations from these estimations. Additionally, the controllers are designed to regulate the velocity, rate of descent and vertical position. A full model, including the nonlinear flight dynamics, controller, carrier deck motion, wind and measurement noise, is constructed numerically and implemented in software. Combining the observer with a proportional-derivative (PD) control, the proposed pitch control shows the better transient characteristics and stronger robustness than a proportional-integral-derivative (PID) controller. The simulations verify that the designed control system can make the aircraft quickly track a time-varying reference despite the existence of nonlinear disturbances and noise.

**Keywords** Carrier landing · Unmanned F/A-18 aircraft · Landing control system · Augmented observer

## 1 Introduction

Carrier landings are considered to be the most challenging routine task of airplane operations, mainly due to the small dimensions of the runway, carrier deck motion and wind disturbances [1]. A schematic diagram of carrier landing is shown in Fig. 1. Due to the complexity of carrier landings, navies must spend a lot of additional resources to rigorously train pilots. Furthermore, the current autopilots are not robust enough to be used for carrier landings.

Recent years, the demands for carrier-based UAVs are increasing. In [2], in addition to introducing the world's first autonomous aerial refuelling system (i.e., MQ-25A Stingray) that is being developed and tested for the long-range F/A-18 E/F fighters, it is emphasized and analysed that making an unmanned vehicle land on a moving runway is the biggest challenge. This highlights that a robust UAV control system capable of landing on an aircraft carrier is important to the development of carrier based UAVs and a very current problem. Highly nonlinear wind disturbances make it difficult for an aircraft to track a reference trajectory because they influence the aerodynamic forces and moments by varying the airspeed. This is especially true for a trajectory that is changing due to the carrier deck motion. Therefore, a method of rejecting the nonlinear disturbance and adjusting the flight path is a crucial element in a robust carrier landing control system. Hence, the focus of this paper investigates whether an augmented observer developed by Wang et al [3] for use in quadrotor aircraft can be used to estimate the nonlinear disturbances and reject noise in the carrier landing context. The disturbance estimations by the observer are used to design a controller for longitudinal pitch dynamics of an unmanned F/A-18 fighter. This controller is tested inside a complete carrier landing system model and verified in simulation.

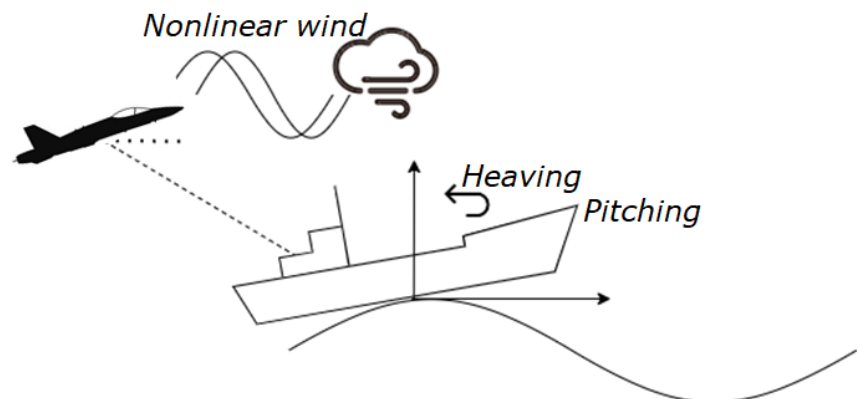

**Fig. 1 Schematic diagram of carrier landing**

## 2 Background

### 2.1 Literature

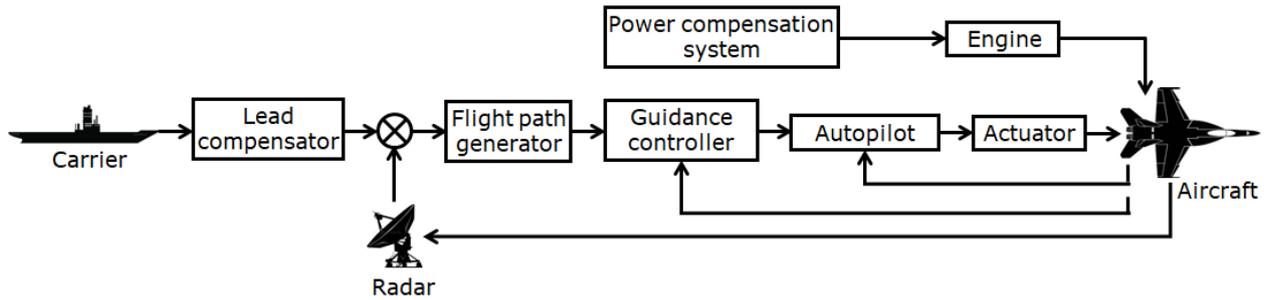

**Fig. 2 Typical carrier landing architecture**

Most of the current research is still based around the original AN/SPN-42 automatic carrier landing system architecture shown in Fig. 2. This system was based on a radar tracking the separation between the aircraft and carrier [4]. For this system, position errors are calculated based on the aircraft's range, altitude and desired glide slope which are filtered and went through a PIDDD guidance controller. This outputs the pitch and roll commands to manoeuvre the aircraft to the required glide path. The commands are implemented through the aircraft's Automatic Flight control system (AFCS) (or autopilot) which tracks the commands by setting the actuator signals. Additionally, in [4], an automatic power compensation system (APCS) was used to maintain a reference angle of attack by controlling the throttle. During most of the approach process, the aircraft is directed to the average touchdown position. Within the last 12 seconds of landing, the aircraft begins to track the exact position of the touchdown position by an additional altitude command to the vertical error determined by using the lead compensation added to the deck position measurement which predicts the future position of the touchdown point [4, 5]. In general, current studies used the same architecture and only focused on improving individual elements.

In [6], the effectiveness of three different guidance controllers, based on PID, PIDDD and Fuzzy PID respectively, were compared for adjusting the flight path. It is shown that the controllers to all be effective when no turbulence is present. However, they all fail when an aerodynamic disturbance is introduced. This highlights the importance of effectively rejecting the aerodynamic disturbance during aircraft landing on moving carrier. The guidance controller is not crucial in mitigating turbulence that this is the function of the autopilot.

### 2.2 Technical problems

The relevant nomenclature is shown in Table 1.

**Table 1 Nomenclature**

| | Name | Symbol | Unit |
|---|---|---|---|
| **Constants** | Aircraft heavy or max trap weight | $m$ | $kg$ |
| | Air density | $\rho$ | $kgm^{-3}$ |
| | Gravity Acceleration | $g$ | $ms^{-2}$ |
| | Maximum thrust | $T_{max}$ | $N$ |
| | Wing area | $S$ | $m^2$ |
| | Aircraft moment inertia around y-axis | $J_y$ | $kgm^2$ |
| | Mean aerodynamic chord | $\overline{c}$ | $m$ |
| **Control Variables** | Airspeed | $V_T$ | $ms^{-1}$ |
| | Inertial speed | $V$ | $ms^{-1}$ |
| | Pitch angle | $\theta$ | $rad$ |
| | Angle of attack | $\alpha$ | $rad$ |
| | Pitch rate | $q$ | $rads^{-1}$ |
| | Flight path angle | $\gamma$ | $rad$ |

| Control inputs | Elevator deflection | $\delta_e$ | ° |
|---|---|---|---|
| | Thrust | $T$ | $N$ |
| | Throttle | $\delta_T$ | - |
| Aerodynamic variables | Lift force | $L$ | $N$ |
| | Drag force | $D$ | $N$ |
| | Pitching moment | $M$ | $N$ |
| | Dynamic pressure | $\bar{q}$ | $Pa$ |
| | Wind speed | $V_w$ | $ms^{-1}$ |

### 2.2.1 Nonlinear flight dynamics

#### 2.2.1.1 Motion equations

The aircraft longitudinal plane is shown in Fig. 3 with the variables, the axes, the forces and the moments.

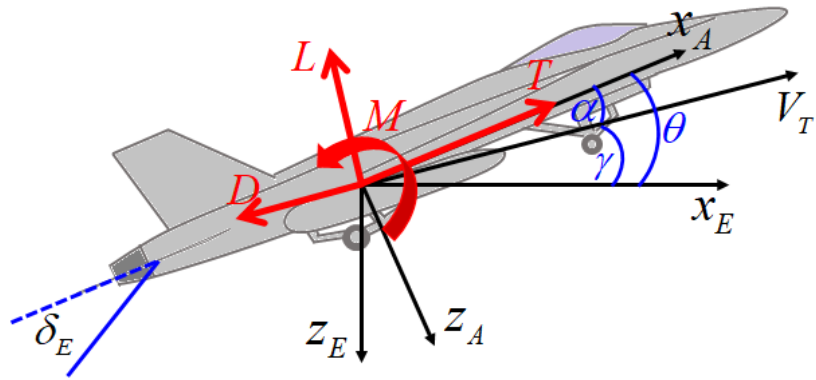

**Fig. 3 Aircraft longitudinal plane**

The longitudinal dynamics are governed by four coupled nonlinear differential equations shown in Eqs. (1)-(4). These equations are described in [7] (Eq. 2.5-32 in Chapter 2, p.115).

$$\dot{V}_T = \frac{T\cos(\alpha)}{m} - \frac{D}{m} - g\sin(\gamma) \tag{1}$$

$$\dot{\theta} = q \tag{2}$$

$$\dot{\alpha} = -\frac{T\sin(\alpha)}{mV_T} - \frac{L}{mV_T} + g\cos(\gamma) \tag{3}$$

$$\dot{q} = \frac{M}{J_y} \tag{4}$$

#### 2.2.1.2 Control inputs

The aircraft has two control inputs in the longitudinal plane, i.e., the thrust and the elevator deflection, and they are provided by: 1) two F404-GE-400 enhanced performance turbofan engines, up to 16,000 lbf thrust each; 2) elevator with its deflection range -25°~10° [8]. The thrust mainly affects the airspeed, especially at small angle of attacks. The elevator mainly affects the pitching moment which induces a pitch rate and hence a change in the pitch angle. Additionally, the elevator deflection causes the minor changes in the drag and lift values.

#### 2.2.1.3 Implications

Eqs. (1)-(4) are nonlinear. Hence, the linear control methodologies cannot be used directly for accurate control. These equations must either decoupled and linearised or nonlinear control methodologies need to be used.

The system expressed by Eqs. (1)-(4) is underactuated, meaning there are some variables that cannot be controlled directly. For example, the vertical position must be controlled by changing the pitch via the elevator which induces a change in the trajectory.

### 2.2.2 Wind disturbance

For aircraft landing on aircraft carrier, many times, wind disturbance (e.g., gust) affects the relative wind and brings the change in angle of attack. Therefore, according to Eqs. (5) and (6), the disturbance effects exist in the aerodynamic forces and moments, affecting the control performance adversely.

$$\bar{q} = 0.5\rho V_T^2 \tag{5}$$
$$L, D, M = \bar{q}SC_{L,D,M}(\alpha) \tag{6}$$

### 2.2.3 Carrier deck motion

Carrier deck motion changes the position of the landing point. Hence a constant trajectory cannot be used for approach. Instead, a sink rate or position commands must be sent to the autopilot to track the carrier deck motion. These movements are caused by the sea waves and can be split into two motions in the longitudinal plane: a heaving (vertical) translational motion of the ship's centre of mass and a pitching rotational motion around its centre of mass. This causes a combined planar motion in the landing position. The aircraft's trajectory must track this motion to land successfully.

### 2.2.4 Measurement noise

Measurement noise is the error between the actual and measured value of a variable. Often, it is in the form of a low-amplitude, high-frequency wave or random stochastic disruption. This introduces the uncertainties into the system and behaves like a disturbance. This makes it more difficult to track a reference trajectory and increases actuator oscillations which dissipates energy.

## 2.3 Nonlinear augmented observer

A nonlinear augmented observer detailed in [3] is used to estimate and compensate for the disturbance effect in the aircraft pitch dynamics. The observers are designed to estimate unknown variables by comparing the error between the measured output and the first variable of the observer. This error is used in a closed feedback loop to improve the accuracy of the estimates. These estimates can then be used in a controller to improve control performance or regulate the unknowns.

This augmented observer is designed to use the perturbation method and is shown to be capable of synchronously estimating the unknown variable and the system disturbance with finite-time stability and stochastic disturbance rejection.

The augmented observer is applied to the system in the form shown below in Eqs. (7)-(10), where, $w_1$, $w_2$ and $w_3$ represent the system variables, $h(t)$ is the known function, $w_3$ is the unknown bounded disturbance with the unknown derivative $\eta(t)$, $y_{op}$ is the measurement output, and $n(t)$ is the measurement noise.

$$\dot{w}_1 = w_2 \tag{7}$$

$$\dot{w}_2 = w_3 + h(t) \tag{8}$$

$$\dot{w}_3 = \eta(t) \tag{9}$$

$$y_{op} = w_1 + n(t) \tag{10}$$

The observer is implemented using the form expressed by Eqs. (11)-(13), where the variables $x_1, x_2$ and $x_3$ represent estimates of $w_1, w_2$ and $w_3$ respectively.

$$\dot{x}_1 = x_2 - \frac{k_3}{\varepsilon}\left|x_1 - y_{op}\right|^{\alpha_3} sign(x_1 - y_{op}) \tag{11}$$

$$\dot{x}_2 = x_3 - \frac{k_2}{\varepsilon^2}\left|x_1 - y_{op}\right|^{\alpha_2} sign(x_1 - y_{op}) \tag{12}$$

$$\dot{x}_{1,3} = -\frac{k_1}{\varepsilon^3}\left|x_1 - y_{op}\right|^{\alpha_3} sign(x_1 - y_{op}) \tag{13}$$

The parameters $k_1, k_2, k_3, \alpha_1, \alpha_2, \alpha_3$ and $\varepsilon$ vary the system stability, precision and robustness and must be selected appropriately. Justifications and rules for selecting these parameters are rigorously explained in [3]. The main conclusions on the observer are summarised below.

- $k_1, k_2, k_3$ is selected such that $\left\{ k_1, k_2, k_3 \mid k_1 > 0, k_3 > 0, \ k_2 > \frac{4}{\pi} \frac{k_1}{k_3} \right\}$ so that the Routh-Hurwitz stability criterion is satisfied, and then the observer is stable.

- Increasing $\alpha_1$ improves the estimation precision, where $\alpha_1 \in (0,1)$, $\alpha_2 = \frac{2\alpha_1 + 1}{3}$ and $\alpha_3 = \frac{\alpha_1 + 2}{3}$.

- Reducing $\varepsilon$ (where, $\varepsilon \in (0,1)$) increases the bandwidth of the observer. Enhancing $\varepsilon$ (where, $\varepsilon \in (0,1)$) decreases the bandwidth of the observer.

## 3 Methodology

This section describes the numerical models used. Additionally, the description of the control method and simulation is provided.

### 3.1 Aircraft model

As mentioned, the longitudinal dynamics of an aircraft are governed by Eqs. (1)-(4). The parameters of the F/A-18 HARV are obtained from [8, 9, 10] and are summarised in Table 2. Due to the low altitude range, the density is assumed to be constant as the airspeed fluctuations caused by the wind disturbance dominate over the minor density changes with altitude.

**Table 2 Constants ([8, 10])**

| Symbol | Value | Unit |
|--------|-------|------|
| $m$ | 15000 | $kg$ |
| $\rho$ | 1.33 | $kgm^{-3}$ |
| $g$ | 9.75 | $ms^{-2}$ |
| $T_{max}$ | 71172 | $N$ |
| $S$ | 37.16 | $m^2$ |
| $J_y$ | 205000 | $kgm^2$ |
| $\bar{c}$ | 3.51 | $m$ |

The aerodynamic forces are the functions of the dynamic pressure and the aerodynamic coefficients $C_L, C_D$ and $C_M$. The aerodynamic coefficients can be approximated as functions of the angle of attack, pitch rate and elevator deflection. These coefficients of the F/A-18 HARV are from [8] and [9], and they are based on experimental data collected during test flights. These coefficients are assumed to be in a clean no-flap configuration. During landing, the flaps of an F/A-18 would be deployed which would change the $C_L$ curve up to some extent. However, the estimations of the designed augmented observer include the effect from this change, and the estimations are used for compensation in control. The coefficients at $\delta_e$ and $q$ set to zero are shown in Fig. 4.

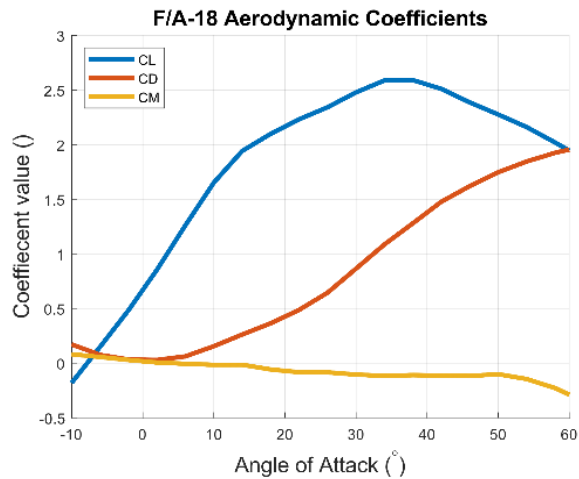

**Fig. 4 F/A-18 Aerodynamic coefficients**

## 3.2 Linearised form

The estimation of the augmented observer (11)-(13) includes all the nonlinearities in the pitch dynamics. Hence, the linear control methods can be used to design the pitch controller. Therefore, the equations of system motion must be linearised around the equilibrium (trim) condition.

### 3.2.1 Equilibrium trim conditions

The trim conditions selected is the steady level flight for the following reasons:

- The aircraft may need to climb or sink to track the ship motion. The level position is the neutral position between climb and sink.

- The trim conditions are simple, and the trim values are easy to obtain.

The trim conditions (denoted by *) for the steady level flight are detailed in [7], and they are well known to be

$$L^* = mg, \qquad T^* = D^*, \qquad M^* = 0$$

Given the partial derivative of the aerodynamic coefficients with respect to the angle of attack $C_{L_\alpha}, C_{D_\alpha}$ and $C_{M_\alpha}$, the trim variables are easy to obtain. However, the model used does not include these partial derivatives. Hence, they must be estimated. Therefore, the trim process shown in Fig. 5 includes an additional step where the aerodynamic coefficients and airspeed are updated to match the estimated values of $\alpha$ and $\delta_e$.

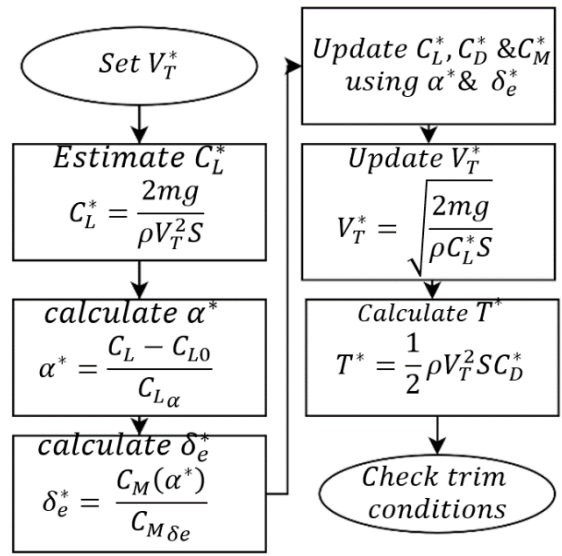

**Fig. 5 Trim flowchart**

After trial and error, the aircraft was trimmed with the parameter shown in Table 3.

**Table 3 Trim variables**

| Trim variable | Value | Unit |
|---|---|---|
| $V_T^*$ | 69.1 | $ms^{-1}$ |
| $\theta^*$ | 7.1 | ° |
| $\alpha^*$ | 7.1 | ° |
| $q^*$ | 0 | $rads^{-1}$ |
| $\gamma^*$ | 0 | $rad$ |

### 3.2.2 Small perturbation equations

Using these trim values, a linear state-space form of the system equations was obtained using a third-order approximation based on the Taylor series expansion in [7] (p.199-202). This approximation is shown in Eq. (14) and is a function of small perturbations (denoted by $\Delta$) from the trim conditions.

$$\begin{bmatrix} \Delta\dot{V}_T \\ \Delta\dot{\theta} \\ \Delta\dot{\alpha} \\ \Delta\dot{q} \end{bmatrix} = \begin{bmatrix} -0.18 & -9.81 & -0.274 & 0 \\ 0 & 0 & 0 & 1 \\ -0.0041 & 0 & -0.59 & 1 \\ 0 & 0 & -0.26 & -0.15 \end{bmatrix} \begin{bmatrix} \Delta V_T \\ \Delta\theta \\ \Delta\alpha \\ \Delta q \end{bmatrix} + \begin{bmatrix} -0.001 & 9.85 \\ 0 & 0 \\ -0.00075 & -0.018 \\ -0.015 & 0 \end{bmatrix} \begin{bmatrix} \Delta\delta_e \\ \Delta\delta_t \end{bmatrix} \tag{14}$$

This approximation is verified by examining the eigenvalues of the A matrix which are $-0.40 \pm 0.45i$ and $0.022 \pm 0.17i$. As expected, these eigenvalues correspond to the two natural modes, i.e., the short-period mode and phugoid mode. Additionally, an open-loop throttle step response is examined, and Fig. 6 shows the very good agreement between the nonlinear and linear models which verifies the approximation.

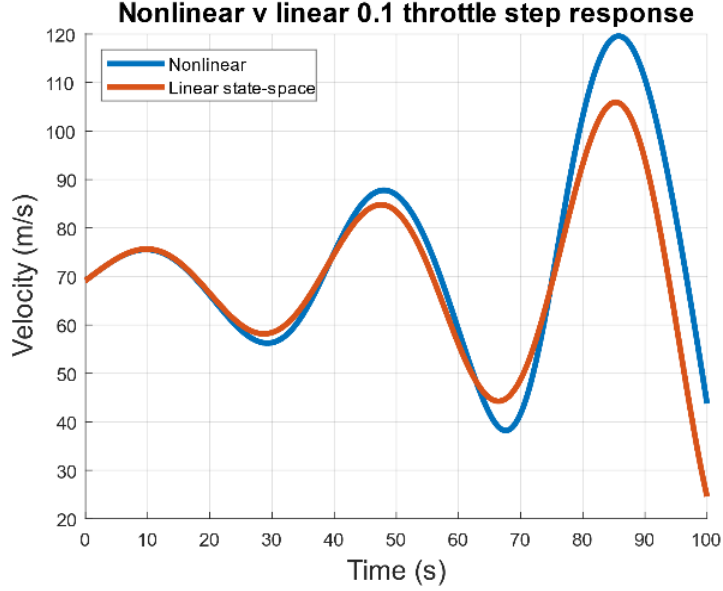

**Fig. 6 Nonlinear v Linear Open-Loop step response**

### 3.3 Ship motion

The heaving and pitching of the ship is modelled by the power spectral density functions $\Phi_H(s)$ and $\Phi_\theta(s)$ from [1]. These are the transfer functions which output the ship motions given a white-noise input. Eqs. (15) and (16) are the ones used here and represent an Essex class carrier.

$$\Phi_H(s) = \frac{1.21}{s^4 + 2.08s^3 + 1.32s^2 + 0.4s + 0.16} \tag{15}$$

$$\Phi_\theta(s) = \frac{0.773s^2}{s^4 + 2.08s^3 + 1.32s^2 + 0.4s + 0.16} \tag{16}$$

In fact, the values of an Essex class can be extrapolated to a modern carrier with the F/A-18 aircrafts. The designed augmented observer doesn't require the specific and precise models of aircrafts and carriers. It means that the designed observer is fit for many types of dynamical systems, and the modelling uncertainties of these systems can be estimated precisely by the observer.

Assuming the landing point $(x_L, z_L)$ to be 81m from the centre of mass coordinates $(x_G, z_G)$ [1], the combined landing point movement can be expressed by:

$$x_L = x_G - 81\cos(\theta_s) \tag{17}$$
$$z_L = z_G - 81\sin(\theta_s) \tag{18}$$

A power was not specified for the white noise in [1]. Hence, the power was tuned to match the maximum pitch and heave oscillation of 3° and 4m detailed in [11]. The power was set to -20dB and 4.5dB for the pitch and heave, respectively. A typical landing point z-displacement can be seen in Fig. 7.

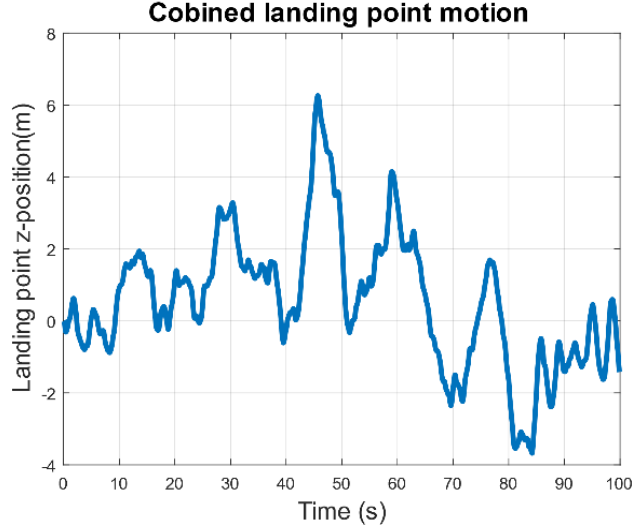

**Fig. 7 Landing point z-motion**

### 3.4 Wind

Based on the model detailed in the military specification [12], the x and z wind disturbances denoted by $u_g$ $w_g$ contain four components: Free-air turbulence components and steady, periodic and random components of the carrier air wake. The free air turbulence $(u_1, w_1)$ components are modelled by white-noise filters $\Phi_{u_1}$, $\Phi_{w_1}$, which are expressed in Eqs. (19) and (20), similar to the power spectral density functions used in Eqs. (15) and (16).

$$\Phi_{u_1} = \frac{200}{1 + (100\Omega)^2} \tag{19}$$

$$\Phi_{w_1} = \frac{71.6}{1 + (100\Omega)^2} \tag{20}$$

The air wake components are all given as the functions of the carrier wind over deck $V_{w/d}$ which is assumed to be $10\ ms^{-1}$ which is in line with the landing axis. The steady components $(u_2, w_2)$ are presented as the profiles. However, they need to be approximated as the linear discontinuous functions of the distance $X$ from the ships centre of pitch dynamics (see Eqs. (21) and (22)).

$$u_2 = \begin{cases} 0.002X, 0 < X < 914m \\ 0, \qquad X < 0 \end{cases} \tag{21}$$

$$w_2 = -1 + 0.0013X, \quad X < 914m \tag{22}$$

The periodic component is modelled as a function of ship pitch amplitude and frequency $\omega_p\ \theta_s$ which are approximated to be $1.25\ rads^{-1}$ and $0.05rad$ respectively based on the outputs of Eqs. (15) and (16). The original formula also includes the ship velocity which is assumed to be $0\ ms^{-1}$. Hence, the periodic components are calculated using Eqs. (23)-(25). The random components are omitted.

$$u_3 = \theta_s V_{w/d}(2.22 + 0.000091X)C \tag{23}$$

$$w_3 = \theta_s V_{w/d}(4.98 + 0.0018X)C \tag{24}$$

$$C = \cos\left(\omega_p\left[2.28t + \frac{X}{0.85V_{w/d}}\right] + 0.1\right) \tag{25}$$

The x-axis wind 400m behind deck is shown in Fig. 8.

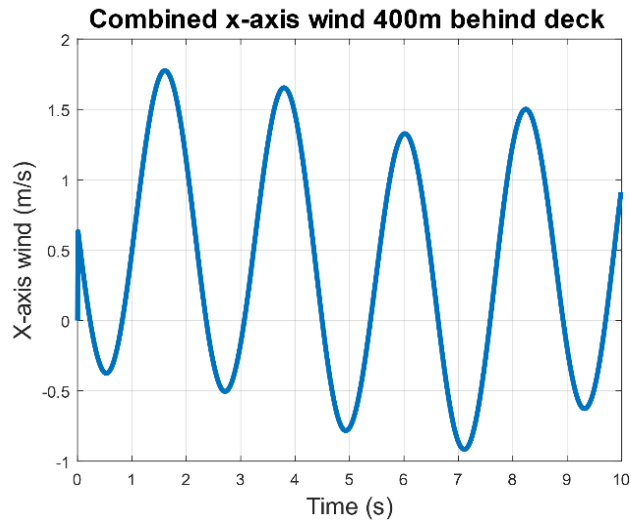

**Fig. 8 x-axis wind 400m behind deck**

## 3.5 Measurement noise

The pitch measurement noise is modelled by the periodic component $0.001\sin(7t)$ and a random component generated by the white noise generator in MATLAB using a power of -60dB.

## 3.6 Actuators

One of the two F/A-18 HARV engine dynamics is approximated by a first-order-lag dynamics on throttle, and the time constants are 0.625 sec and 0.55 sec for non-A/B and A/B, respectively [13] (Chapter 4.1, p.38). We select the time constant 0.625 for the worse condition. The transfer function for the engine dynamics is expressed by:

$$\text{Engine: } \frac{1}{0.625s + 1};$$

The transfer function for the elevator dynamics is expressed by [13] (Chapter 6.2, p.91):

$$\text{Elevator: } \frac{30.74^2}{s^2 + 2(0.509)(30.74)s + 30.74^2}$$

## 3.7 Simulation

The system models are setup as the systems of first-order differential equations with the known initial conditions to form an initial-value problem. Therefore, a Runge-Kutta 4th order algorithm is used to solve the system numerically. This algorithm is implemented in MATLAB. A time step of 0.001s is used.

# 4 Control Architecture

## 4.1 Overview

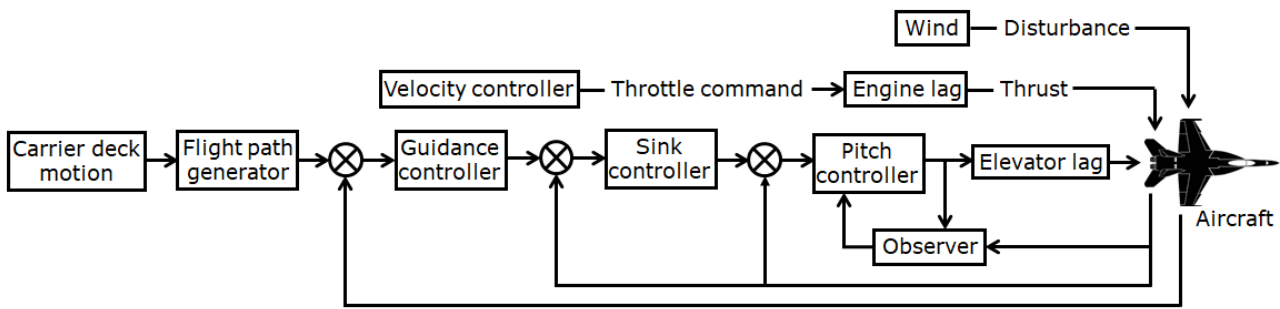

**Fig. 9 Control architecture**

As shown in Fig. 9, the control system architecture used is based on the original architecture described in section 2.1. A flight path generator projects an ideal flight path from the landing position. A position error between the ideal flight path and aircraft is determined and fed into a guidance controller which outputs a required sink rate. This sink rate is fed into a sink controller which outputs a required pitch angle. This command is executed by the pitch controller which sends a control signal to the elevators. Meanwhile, the velocity controller maintains a constant airspeed. The augmented observer uses the information from the outputs of aircraft dynamics and the pitch controller, to estimate the disturbance and improve control performance.

## 4.2 Pitch controller

The pitch controller is designed to track a reference pitch command by controlling the elevator. By observing the disruptions and nonlinearities in a combined disturbance, a linear controller is used to achieve the desired performance.

### 4.2.1 Estimation by observer

Based on Eqs. (2) and (4) and the linearisation form (14), the motion equation for the pitch dynamics is given by Eq. (26).

$$\Delta\ddot{\theta} = \frac{\partial\dot{q}}{\partial q}\Delta\dot{\theta} + \frac{\partial\dot{q}}{\partial\delta_e}\Delta\delta_e + \frac{\partial\dot{q}}{\partial\alpha}\Delta\alpha \tag{26}$$

Equation (26) can be rearranged in terms of the known input $h_\theta(t)$ and the unknown disturbance $d_\theta(t)$ with the derivative $\eta_\theta(t)$ (see Eqs. (27) and (28)). Where, $d_\theta(t)$ represents all the combined disturbances, from the neglected terms, the noise and the nonlinear dynamics omitted in the linear approximation.

$$\Delta\ddot{\theta} = h_\theta(t) + d_\theta(t) \tag{27}$$

$$h_\theta(t) = \frac{\partial\dot{q}}{\partial q}\Delta\dot{\theta} + \frac{\partial\dot{q}}{\partial\delta_e}\Delta\delta_e \tag{28}$$

Therefore, the Eqs. (27) and (28) are in an appropriate form required by the observer and detailed by Eqs. (7)-(10) with $y_{op} = \Delta\theta$. Hence, the observer expressed by Eqs. (11)-(13) can be used to estimate the pitch angle perturbation, pitch rate and disturbance denoted by $\widehat{\Delta\theta}, \widehat{\Delta q}, \widehat{d_\theta}(t)$ with $x_1, x_2, x_3$ respectively.

### 4.2.2 Controller

By defining the errors as

$$e_\theta = \theta_r - \theta = \Delta\theta_r - \Delta\theta \,, \dot{e}_\theta = -\Delta\dot{\theta} \; \ddot{e}_\theta = -\Delta\ddot{\theta}$$

and using Eqs. (26), (27) and (28), the error system can be written by

$$\ddot{e}_\theta - \frac{\partial\dot{q}}{\partial q}\dot{e}_\theta = -\frac{\partial\dot{q}}{\partial\delta_e}\Delta\delta_e - d_\theta(t) = -U_{e\theta}(t) \tag{29}$$

Where, $U_{e\theta}(t)$ is an equivalent controller selected as a proportional-derivative controller

$$U_{e\theta}(t) = K_{P\theta}e_\theta(t) + K_{D\theta}\dot{e}_\theta \tag{30}$$

Hence, the characteristic equation can be expressed by

$$s^2 + \left(K_{D\theta} + \frac{\partial\dot{q}}{\partial q}\right)s + K_{P\theta} = 0 \tag{31}$$

This second-order error system can be expressed in terms of natural frequency $\omega_n$ and damping ratio $\gamma_1$ which can be selected to achieve the desired system performance:

$$s^2 + 2\gamma_1\omega_n s + \omega_n^2 = 0 \tag{32}$$

$$K_{IV} = K_{P\theta}, \; K_{D\theta} = 2\gamma_1\omega_n - \frac{\partial\dot{q}}{\partial q} \tag{33}$$

The natural frequency can be expressed in terms of the convergence time to reach 2% of the required value $T_{2\%}$, as follows:

$$\omega_n = \frac{4}{T_{2\%}\gamma} \tag{34}$$

The damping ratio is selected as $\sqrt{2}$ to speed up the convergence. The desired settling time is reduced to 0.3s until there is no improvement in performance shown by the simulated convergence time. The controller parameters are selected as:

$$K_{D\theta} = 26.5186, \qquad K_{P\theta} = 88.89$$

Although the disturbance is unmeasurable, the observer estimation can be used instead making all variables in the equation known. Therefore, the equivalent controller and disturbance estimation can be used to determine the change in the elevator deflection as follows:

$$\Delta\delta_e = \frac{U_{e\theta}(t) - \widehat{d_\theta}(t)}{\frac{\partial \dot{q}}{\partial \delta_e}} \tag{35}$$

By summing the elevator perturbation with the trimmed value, the final elevator deflection can be determined as:

$$\delta_e = \delta_e^* + \Delta\delta_e \tag{36}$$

### 4.2.3 PID control for comparison

A PID pitch controller is also implemented for comparison with the presented method. The PID equivalent controller can be expressed as

$$U_{e\theta 2}(t) = K_{P\theta 2}e_\theta(t) + K_{I\theta}\int_0^t e_s(\tau)d\tau + K_{D\theta 2}\dot{e}_\theta \tag{37}$$

with the controller parameters shown below:

$$K_{P\theta 2} = 57.01, \qquad K_{I\theta} = 50, \qquad K_{D\theta 2} = 17.19$$

## 4.3 Velocity controller

The velocity controller is designed to maintain a constant velocity by controlling the engine thrust. Due to the relatively low coupling between the airspeed and the other parameters, the velocity controller can be designed directly. Define the error by assuming a constant reference, as follows:

$$e_V = V_R - V_T, \qquad \dot{e}_V = -\dot{V}_T$$

Based on Eq. (1), the error system can be expressed by:

$$\dot{e}_V = \dot{V}_R - \frac{T\cos(\alpha)}{m} + \frac{D}{m} + g\sin(\gamma) \tag{38}$$

By assuming the small angles and grouping the other state variables into a combined disturbance $d_v$, the error system becomes:

$$\dot{e}_V = -\frac{T}{m} + d_V(t) = -U_{eV}(t) \tag{39}$$

Selecting the equivalent controller such that the error system becomes an autonomously stable system with the ability of rejecting unknown constant disturbance [14] (Lemma 2.1, p. 2487). A PID controller is selected as follows:

$$U_{eV}(t) = K_{DV}\dot{e}_v(t) + K_{PV}e_v(t) + K_{IV}\int_0^t e_v(\tau)d\tau \tag{40}$$

Hence, the characteristic equation is

$$(1 + K_{Dv})s^2 + K_{PV}s + K_{IV} = 0 \tag{41}$$

Once the equivalent controller is calculated from the error terms, the actual thrust command can be determined as follows:

$$T = m\left(U_{eV}(t)\right) \tag{42}$$

## 4.4 Sink controller

The sink controller is responsible for controlling the aircrafts rate of descent (sink rate) $\dot{z}$ by inputting a sink rate error $e_s$ and outputting a required pitch angle $\theta_R$. Due to the coupled nature of the motion equations, a fixed relationship between the pitch and sink rate is difficult to obtain as it also depends on the other state variables. Therefore, a proportional-integral (PI) controller is applied directly and tuned manually until an acceptable performance is achieved. Defining the error as $e_s = \dot{z}_R - \dot{z}$, the sink controller can be expressed by

$$\theta_R(t) = K_{Ps}e_s + K_{Is}\int_0^t e_s(\tau)d\tau \tag{43}$$

Where, the controller parameters is selected as

$$K_{PS} = 0.0061, \ K_{IS} = 0.018$$

## 4.5 Guidance controller

The guidance controller takes the vertical deviation $e_z$ and outputs a sink rate command $\dot{z}_R$. Similarly, to the sink rate controller, a PID controller is applied directly and tuned manually. Defining the error as $e_z = z_R - z$, the controller can be expressed by:

$$\dot{z}_R(t) = K_{Pz}e_S + K_{Iz} \int_0^t e_z(\tau)d\tau + K_{Dz}\dot{e}_z(t) \tag{44}$$

Where, the controller parameters are selected as:

$$K_{Pz} = 1, \qquad K_{Iz} = 0.5, \qquad K_{Dz} = 0.01$$

# 5 RESULTS

Figures 10, 11 and 12 show the performance of the observer-based Proportional-Derivative (O-PD) controller vs the PID controller. In the near ideal conditions, where there is no additional noise or wind disturbance, both controllers converge to the required value under 3 seconds.

As shown in Fig. 10, although the steady-state (s-s) error is under 1% for both controllers, the pitch angle by the O-PD controller converges in 1.11s, 85% faster than the one by the PID controller, and the better transient process is performed. In the adverse case with both noise and wind present as shown in Fig. 11, the PID controller completely fails to stabilise the pitch dynamics, whereas the O-PD controller makes the pitch angle converges to within 2% of the required value within 1.4s which is 26% slower than in the ideal case. In Fig. 12, the jagged shape and noise in the elevator deflection highlights the existence of nonlinearity and disturbance in the system. Despite this, the O-PD controller still manages to quickly stabilise the system. Figure 13 shows the sink rate error and trajectory of the aircraft when tracking a constant sink rate of $10 \ ms^{-1}$. The sink-controller to O-PD controller is shown to be adequate, converging to the required value within 0.5s with the negligible steady-state error.

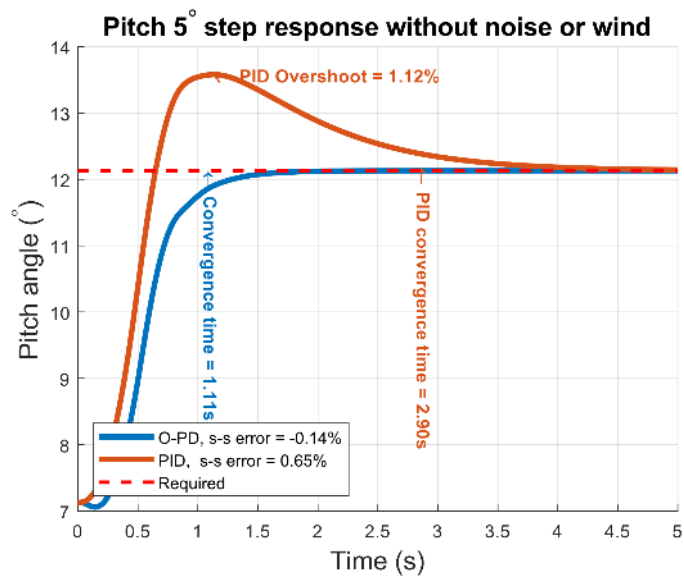

**Fig. 10 Pitch step response without noise or wind disturbance**

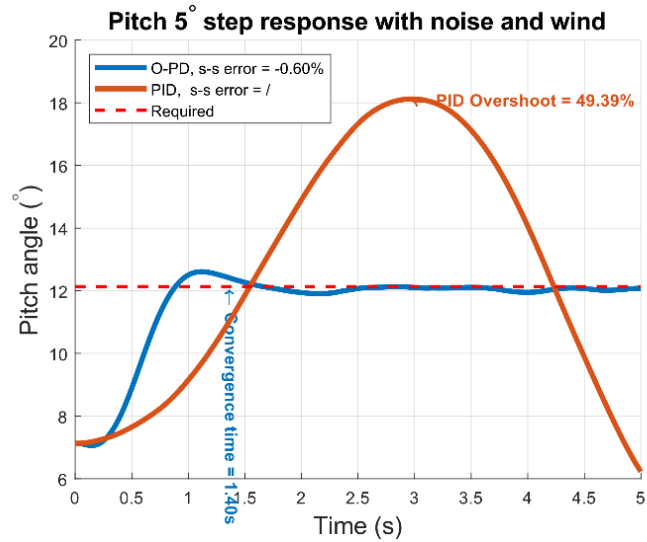

**Fig. 11 Pitch step response with noise and wind disturbance**

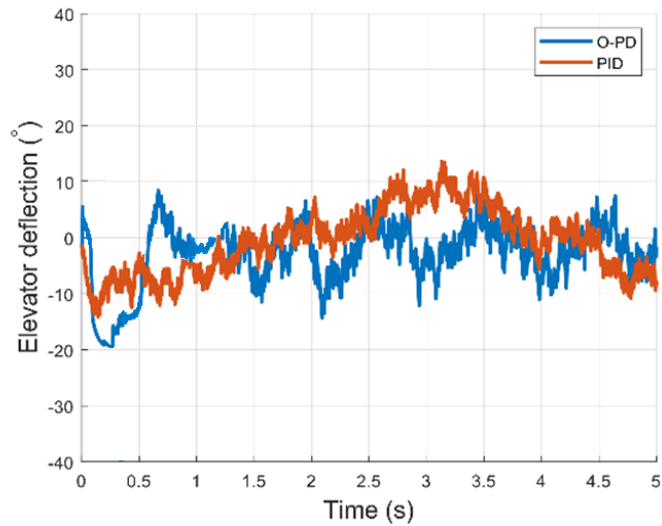

**Fig. 12 Elevator deflection during step response with noise and wind disturbance**

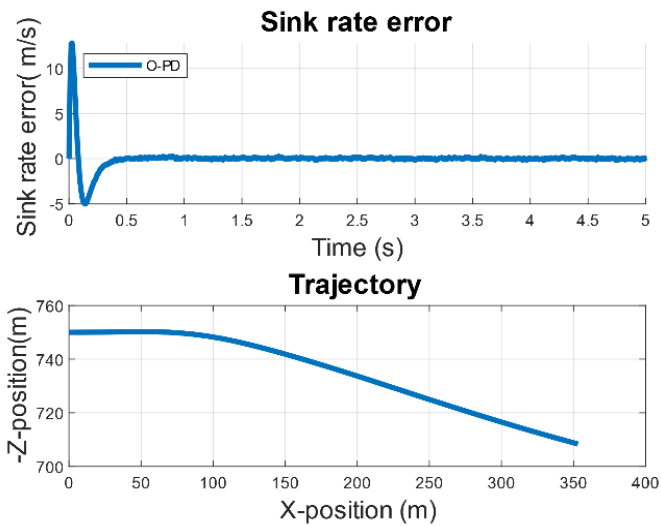

**Fig. 13 Sink rate error and trajectory for a constant required sink rate**

The position error between the aircraft and the ideal glide path is shown in Fig. 14. The sink and pitch controllers are shown to perform well. Figure 15 shows that the pitch angle by the O-PD controller tracks the highly time-varying required value precisely. This highlights the efficient performance of the O-PD controller.

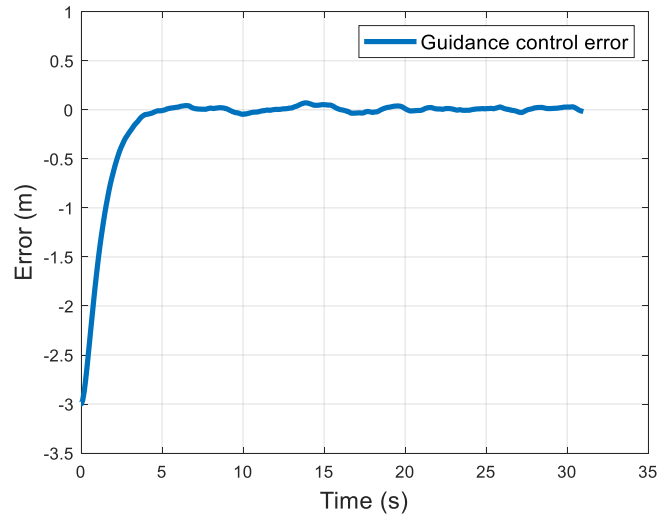

**Fig. 14 Position error with respect to the ideal glide path**

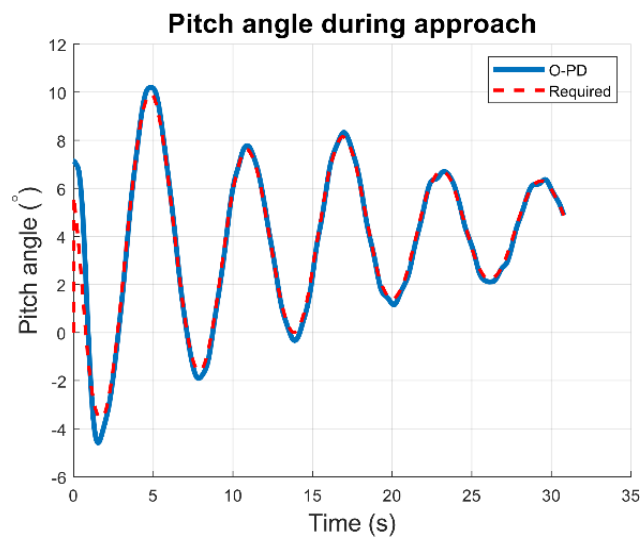

**Fig. 15 Pitch angle response on approach**

## 6 CONCLUSION

The existence of adverse conditions, i.e., nonlinear flight dynamics, strong nonlinear wind disturbance, and time-varying reference trajectory, affect adversely the control performance for carrier landing. A numerical model of an F/A-18 aircraft with the adverse conditions has been developed and implemented. In the presented observer-based control system, a nonlinear augmented observer is applied to estimate the combined disturbances in the aircraft pitch dynamics. Together with a PD controller, the estimate is used for control compensation to achieve satisfactory pitch control performance. The controller is shown to make the pitch angle converge 85% faster than a PID controller, and the better transient process is achieved. Importantly, the presented control method is appropriate for many nonlinear dynamical systems with time-varying reference in the presence of (highly) nonlinear disturbances. Our further work will focus on applying the O-PD controller to the roll and yaw dynamics.